\documentclass[amsfonts,amssymb,amsmath,prl,aps,reprint,notitlepage,longbibliography]{revtex4-2}%

\pdfoutput=1

\usepackage[T1]{fontenc}
\usepackage{times}
\usepackage{xcolor}
\usepackage[colorlinks,allcolors=blue]{hyperref}
\usepackage{graphicx}
\usepackage{bm}
\usepackage{mathrsfs}
\usepackage{dcolumn}
\usepackage{orcidlink}

\usepackage[normalem]{ulem}
\usepackage{dsfont}

\renewcommand{\vec}[1]{{\ensuremath{\bm{\mathrm{#1}}}}}
\newcommand{\uvec}[1]{{\ensuremath{\hat{\bm{\mathrm{#1}}}}}}

\newcommand{\rexp}[1]{\ensuremath{{\mathrm{exp}{\left(#1\right)}}}}

\newcommand{\rmd}{{\ensuremath{\mathrm{d}}}}
\newcommand{\kB}{{\ensuremath{k_{\mathrm{B}}}}}

\begin{document}

\title{Metadynamics calculations of the effect of thermal spin fluctuations on skyrmion stability}

% \author{Ioannis Charalampidis}
\author{Ioannis Charalampidis\,\orcidlink{0009-0007-1174-5883}}
\email{ioannis.charalampidis@manchester.ac.uk}
\affiliation{School of Physics and Astronomy, University of Leeds, Leeds LS2 9JT, United Kingdom}

% \author{Joseph Barker}
\author{Joseph Barker\,\orcidlink{0000-0003-4843-5516}}
\email{j.barker@leeds.ac.uk}
\affiliation{School of Physics and Astronomy, University of Leeds, Leeds LS2 9JT, United Kingdom}

\begin{abstract}
    The stability of magnetic skyrmions has been investigated in the past, but mostly in the absence of thermal fluctuations. However, thermal spin fluctuations modify the magnetic properties (exchange stiffness, Dzyaloshinskii--Moriya interaction (DMI) and anisotropy) that define skyrmion stability. Thermal magnons also excite internal skrymion dynamics, deforming the skyrmion shape. Entropy has also been shown to modify skyrmion lifetimes in experiments, but is absent or approximated in previous studies. Here we use metadynamics to calculate the free energy surface of a magnetic thin film in terms of the topological charge and magnetization. We identify the free energy minima corresponding to different spin textures and the lowest energy paths between the ferromagnetic and single skyrmion states. We show that at low temperatures the lowest free energy barrier is a skyrmion collapse process. However, this energy barrier increases with temperature. An alternative path, where a singularity forms on the skrymion edge, has a larger free energy barrier at low temperatures but decreases with increasing temperature and eventually becomes the lowest energy barrier. 
\end{abstract}

\maketitle

Magnetic skyrmions are particle-like spin textures that may be used to store or transport information~\cite{Finocchio_JPhysDApplPhys_49_423001_2016,Fert_NatRevMater_2_17031_2017,Bogdanov_NatRevPhys_2_492_2020}. Skyrmions are said to have `topological protection', meaning that there is no smooth way to unwind the spin texture back to a uniform state~\cite{Nagaosa_NatNanotechnol_8_899_2013}. Creating or destroying a skyrmion therefore requires a high-energy process to cause a sharp break in the texture, working against the strong exchange interaction. Skyrmions can therefore be created, for example, by applying energy through heating from electric currents~\cite{Yuan_SciRep_6_22638_2016} or lasers~\cite{Gerlinger_ApplPhysLett_118_2021}. Geometric defects such as notches can reduce the energy cost of creating them~\cite{Iwasaki_NatNanotechnol_8_742_2013}. Understanding the energy required to create or destroy a skyrmion is important for both fundamental science and practical applications. So far, calculations of energy barriers for skyrmion creation and annihilation have focused on the nudged elastic band method~\cite{Bessarab_ComputPhysCommun_196_335_2015, CortesOrtuno_SciRep_7_4060_2017}. This method relaxes a series of states (spin configurations) to a saddle point to identify the energy barrier between the initial and final states. To find higher-energy paths, a different choice of the set of states is needed, but convergence to a specific path is not guaranteed. The transition path is also described in terms of an abstract coordinate, which has no physical interpretation, limiting the understanding of the physical process. Most importantly, with respect to our work, nudged elastic band calculations do not account for temperature and, therefore, do not include the role of entropy or the effect of thermal spin fluctuations. Thermal spin fluctuations change magnetic properties such as exchange stiffness, DMI, and anisotropy, all of which determine the properties of skyrmions and even whether skyrmions can exist. Thermal magnons also excite the internal modes of the skyrmion~\cite{Petrova_PhysRevB_84_214433_2011, Mochizuki_PhysRevLett_108_017601_2012, Lin_PhysRevB_89_024415_2014, Kim_PhysRevB_90_064410_2014}. Internal modes may have an impact on the stability and lifetime of the skyrmions. Theoretical and experimental studies have also shown that entropy plays an important role in skyrmion stability and has a significant impact on skyrmion lifetimes~\cite{Wild_SciAdv_3_2017, Desplat_PhysRevB_98_134407_2018}. But nudged elastic band calculations provide only the zero temperature internal energy, $U(0)$, not the free energy $F(T)=U(T)-TS$. The entropy contribution has been included using Langer's theory~\cite{Desplat_PhysRevB_98_134407_2018} but this also requires several approximations and still ignores the renormalization of magnetic parameters.

In this Letter, we use metadynamics, a general numerical method for calculating free energies most commonly used within molecular dynamics~\cite{Laio_ProcNatlAcadSci_99_12562_2002, Barducci_WIREsComputMolSci_1_826_2011}. It has recently been used for spin models to study the temperature dependence of magnetic anisotropy energy and the associated spin reorientation transitions~\cite{Nagyfalusi_JPhysConfSer_903_012016_2017, Nagyfalusi_PhysRevB_100_174429_2019, Nagyfalusi_PhysRevB_102_134413_2020} and within the micromagnetism formalism it has been used to sample the free energy landscape of a vortex in a nanodot, although without studying any thermal effects~\cite{Tobik_PhysRevB_96_140413_2017}. Here, we apply metadynamics with atomistic spin modeling to calculate the free energy landscape of topological spin textures in a 2D lattice. This enables us to calculate the free energy barriers between the ferromagnetic ground state and topological spin textures such as skyrmions. It also gives a complete picture of the energy landscape including alternative, higher-energy paths, as well as identifying other metastable and even unstable spin textures. 

At low temperatures, we find that the minimum energy path for skyrmion annihilation (creation) is through the skyrmion `collapse' process, where the skyrmion shrinks (expands) about a single point~\cite{CortesOrtuno_SciRep_7_4060_2017}. The path with the second-lowest energy barrier corresponds to skyrmion annihilation (creation) through a Bloch singularity on the edge of the skyrmion (domain). Although both of these paths have been identified previously, the energy barrier for the collapse process was always significantly lower than that of the singularity. In this work, we find the two barriers have an opposite temperature dependence, with the energy barrier for the collapse process increasing and the energy barrier for the singularity process decreasing as the temperature increases. There is a crossover temperature at which the singularity path becomes the minimum energy path, 36~K in for our model parameters. We postulate that the configurational entropy plays an important role. Our results demonstrate both how metadynamics can be applied to study topological spin textures and the important role spin fluctuations have on skyrmion stability.

\textit{Method} -- We model a hexagonal lattice with crystal space group P6/mmm (191) using basis vectors $\vec{a}_1 = (1,0,0)$, $\vec{a}_2 = (1/2,\sqrt{3}/2,0)$, $\vec{a}_3 = (0, 0, 1)$ in a $65\times65\times1$ supercell. Periodic boundaries are applied in the $(\vec{a}_1, \vec{a}_2)$ plane, and open boundaries along $\vec{a}_3$, making a 2D system. The Hamiltonian is
\begin{equation}
    \mathcal{H} = -\frac{1}{2}\sum_{\langle ij \rangle} J_{ij} \vec{S}_i \cdot \vec{S}_j 
    - \frac{1}{2}\sum_{\langle ij \rangle} \vec{D}_{ij} \cdot (\vec{S}_i \times \vec{S}_j)
        - \sum_i k_z S_{z,i}^2,
\end{equation}
where $\langle ij \rangle$ indicates that the sum is taken only over the six nearest neighbors. $\vec{S}_i$ are classical spin vectors of length $|\vec{S}_i|=1$, $J_{ij}=13.0~\mathrm{meV}$ are the nearest-neighbor exchange interaction energies, $\vec{D}_{ij} = D_{ij} (\uvec{z}\times\uvec{r}_{ij})$ are the nearest-neighbor DMI interaction vectors where $\uvec{z}$ is a unit vector in the $z$ (out-of-plane) direction, $\uvec{r}_{ij}$ is the unit vector pointing from site $i$ to site $j$ and $D_{ij} = 0.811~\mathrm{meV}$. $k_z=0.09~\mathrm{meV}$ is a uniaxial anisotropy energy. The factors of $1/2$ account for the double counting in the sums. This is a typical Hamiltonian, where for certain combinations of exchange, uniaxial anisotropy and DMI, isolated skyrmions are metastable states even without an applied field. 

We simulate the spin model at finite temperature using the Metropolis Monte Carlo method~\cite{Landau_Binder_Book}. Our Monte Carlo trial moves are small angle changes of a spin from its initial direction: $\vec{S}_i^{\mathrm{trial}} = \vec{S}_i^{\mathrm{initial}} + \varphi\vec{\Omega}$ where $\vec{\Omega}$ is a unit vector uniformly random on the 2-sphere and $\varphi=0.3$. Metadynamics adds a fictitious potential $V$ to the energy of the system, forcing it away from states that have already been sampled, thus exploring new states~\cite{Laio_RepProgPhys_71_126601_2008}. This enables the system to escape energy minima even over high barriers. The probability of accepting a new state $s_b$ from a state $s_a$ differs from the plain Metropolis probability in that it now contains the bias potential
\begin{equation}
    P(s_{a} \rightarrow s_{b}) = \mathrm{min}\left[1, \rexp{-\frac{\Delta E_{ab} + \Delta V_{ab}}{\kB T}} \right]
\end{equation}
where $\Delta E_{ab} = E(s_b) - E(s_a)$ is the difference in total energy calculated from the Hamiltonian and $\Delta V_{ab} = V(s_b) - V(s_a)$ is the difference in the metadynamics potential for states $s_b$ and $s_a$.

Free energy surfaces are generally described with macroscopic coordinates of a system. In metadynamics these are called \textit{collective variables} (CV) and represent macroscopic quantities calculated from microscopic degrees of freedom, the spin vectors in our case. For this study our CVs are the topological charge
\begin{equation}
    Q = \frac{1}{4\pi} \int \rmd^2 x ~\vec{S}\cdot \left(\frac{\partial \vec{S}}{\partial x} \times \frac{\partial \vec{S}}{\partial y}\right),
    \label{eq:topological_charge}
\end{equation}
and the $z$-component of the reduced magnetization 
\begin{equation}
    m_z = \frac{1}{N}\sum_{i}^N S_{i}^{z}.
\end{equation}
where $N$ is the number of spins in the system.

To use the topological charge as a CV to sample different spin textures, we considered several key points:

1) CVs must be continuous variables so that the transitions between states can be sampled smoothly. The skyrmion topological charge should strictly be a discrete integer. However, the calculation of $Q$ as defined in Eq.~\eqref{eq:topological_charge} on a discrete lattice and with spin fluctuations breaks the assumptions of a smooth continuum system and numerical calculations of $Q$ therefore do produce non-integer values~\cite{Kim_IOPSciNotes_1_025211_2020,Note1}. This allows us to use $Q$ as a CV. Other strategies for making topological charges continuous have been used in lattice quantum chromodynamics calculations with metadynamics~\cite{Laio_JHighEnergyPhys_2016_89_2016, Bonanno_JHighEnergyPhys_2019_3_2019}.

\footnotetext[1]{Using the `geometrical definition' of topological charge as a summation over plaquettes~\cite{Berg_NuclPhysB_190_412_1981} always produces an integer and therefore cannot be used as a CV for metadynamics.}

2) The topological charge alone is not suitable for sampling the free energy landscape because multiple spin textures with the same charge, but different energy, cannot be distinguished. For example, with $Q=0$ the system can be single domain or multi-domain and these have different energies. This means that using $Q$ alone as a CV would give an incorrect projection of the energy landscape~\cite{Bussi_NatRevPhys_2_200_2020}. Hence, we use $m_z$ as the second CV to distinguish between different states with the same $Q$.

3) The variable $m_z \in [-1,1]$ is bounded, giving definite limits to the state space which can be explored in the simulation. $Q$, however, is unbounded, meaning that metadynamics can explore ever larger values of $Q$, generating more and more topological charge within our finite lattice. To avoid the system spending a long time exploring high-$Q$ states with many skyrmions we implemented boundary conditions on the metadynamics potential which behave like a harmonic spring, smoothly pushing the system back towards a defined region of interest (see Supplementary Information for more details).

The metadynamics potential $V(Q,m_z)$ is built by adding a Gaussian every full Monte Carlo sweep, $\tau$ (defined as $N$ trial moves). The potential at a given time $t$ is then
\begin{equation}
\begin{aligned}
    V(Q, m_z, t) = \omega_0 \sum_{\substack{t' = \tau, 2\tau, \cdots\\ t' < t}} \rexp{-\frac{V(Q,m_z,t')}{\kB \Delta T}} \\
    \times \rexp{-\frac{(Q(t')-Q(t))^2}{2\sigma_{Q}^2} - \frac{(m_z(t') - m_z(t))^2}{2\sigma_{m_z}^2}}
\end{aligned}
\label{eq:potential}
\end{equation}
where $\omega_0=0.1$~meV is the initial Gaussian amplitude in units of energy and $\sigma_{Q}=0.1$ and $\sigma_{m_z}=0.05$ are widths in the dimensions of $Q$ and $m_z$. The exponential decay in \eqref{eq:potential} is because we have implemented \textit{well tempered} metadynamics \cite{Barducci_PhysRevLett_100_020603_2008} which systematically reduces the Gaussian amplitude to ensure convergence. The tempering bias temperature is $\Delta T = 2500$~K. The free energy is calculated from the metadynamics potential as
\begin{equation}
    F(Q, m_z) = -\frac{T+\Delta T}{\Delta T}V(Q, m_z).
\end{equation}

In a large system, the energy to create and destroy two skyrmions separated by a large distance should be almost independent. However, the size of our simulations is quite restricted (due to computational costs) and the creation energy of a second skyrmion will be influenced by the existence of the first. We have attempted to make the simulation large enough to avoid a significant self-interaction of one skyrmion with itself across the periodic boundaries, but once two skyrmions exist in the system, the energetics of that state and the energy to create or destroy the second skyrmion are likely to be affected. Inserting further skyrmions will further increase the interaction energies. Therefore, we use the harmonic spring boundary conditions to restrict the system to $Q \in [-2.5,2.5]$. In principle calculations could be performed in open systems where the edges provide another vector for skyrmion creation and annihilation, but we have not considered this here. 

We solved $1.9\times10^8$ Monte Carlo steps and ran 10 independent simulations for each temperature, averaging the results to produce the energy surfaces. The standard error at all points on the calculated free energy surface is $<40$~meV.

\begin{figure}
    \centering
    \includegraphics[width=0.48\textwidth]{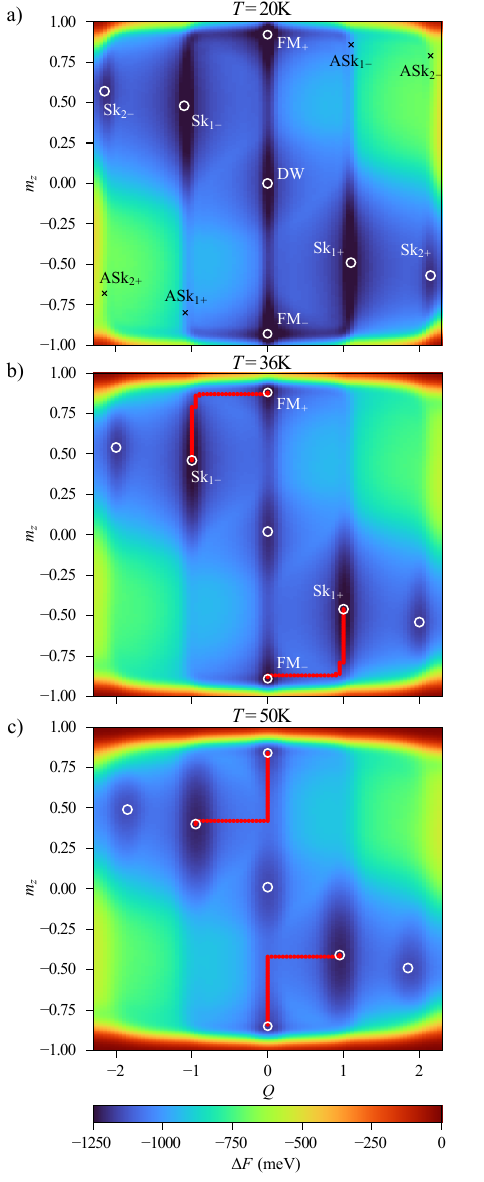}
    \caption{Metadynamics calculated free energy surfaces at $T=20$K, $36$K and $50$K. $\circ$ mark local energy minima $\times$ mark unstable points where there is no minima but a non-trivial spin texture forms. Red dotted lines show the minimum energy path between FM and Sk$_1$ states.}
    \label{fig:free_energy_surfaces}
\end{figure}

\begin{figure}
    \centering
    \includegraphics[width=0.48\textwidth]{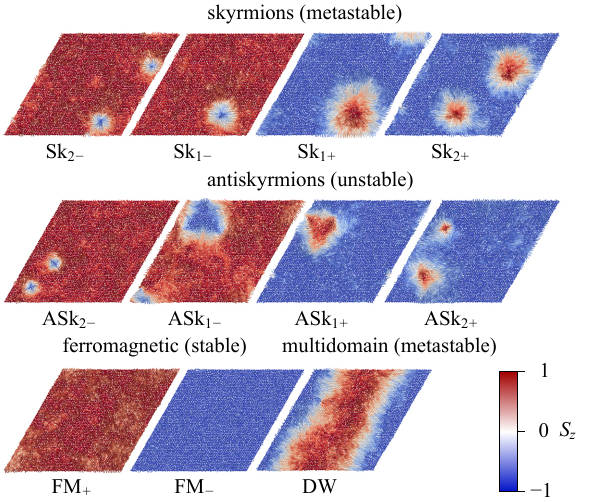}
    \caption{Snapshots of example spin states at different points on the free energy surface as labeled in Fig.~\ref{fig:free_energy_surfaces}. Note that there are periodic boundaries in the plane.}
    \label{fig:skrymion_states}
\end{figure}

\textit{Results} -- Fig.~\ref{fig:free_energy_surfaces}a shows the free energy landscape calculated at $T=20$, $36$ and $50$~K. The Curie temperature is $T_{C} = 136$~K. We mark local energy minima on the figure with white circles. Sample spin configurations from the simulation at the marked points are shown in Fig.~\ref{fig:skrymion_states}. The top and bottom edges of the free energy surface are dark red due to the large free energy cost of states with a magnetization larger than the thermal equilibrium value; Increasing the magnetization against the thermal spin fluctuations requires work against entropy. At $Q=0$ there are three energy minima: $\mathrm{FM}_+$ and $\mathrm{FM}_-$ are global minima corresponding to the uniform ferromagnetic states with magnetization aligned along $+z$ and $-z$ respectively. The point marked $\mathrm{DW}$ at $(Q=0, m_z=0)$  is a two-domain state with half of the magnetization aligned along $+z$ and half along $-z$ with domain walls between. The DMI in the Hamiltonian means that the magnetization does not reverse from $\mathrm{FM}_+$ to $\mathrm{FM}_-$ by a coherent rotation. Instead, domain-wall nucleation and propagation is energetically cheaper. Removing the DMI from the Hamiltonian produces free energy surfaces without the $\mathrm{DW}$ state, showing the expected coherent reversal. This is a significant benefit of metadynamics over the nudged elastic band method in that it is free of assumptions about the initial, final, or transition states. This may be useful in the study of more complex topological magnetic textures such as hopfions~\cite{Rybakov_APLMater_10_2022}.

At $Q=\pm1$ the labeled states $\mathrm{Sk}_{1\pm}$ are single skyrmions with a core in the $\pm z$ direction. We note the whole energy surface has a $\pi$ rotational symmetry; changing the ferromagnetic background from $m_z \rightarrow -m_z$ the sign of $Q$ for which skyrmions are stable swaps. The energy surface is quite shallow between $\pm0.25 \lesssim m_z \lesssim \pm0.7$, showing that the skyrmions are quite soft; there is little energy cost to expand or contract the size of the skyrmion around the equilibrium value. At $Q=\pm 2$ we find minima $\mathrm{Sk}_{2\pm}$ corresponding to states with two skyrmions. In the opposite topological sector from the skyrmions we find anti-skyrmions $\mathrm{ASk}_{1\pm}$ and $\mathrm{ASk}_{2\pm}$, marked with crosses. The DMI vectors in our Hamiltonian are incompatible with forming metastable anti-skyrmions; nevertheless, these states are explored by metadynamics as valid combinations of $Q$ and $m_z$. Although anti-skyrmions are formed, the free energy has no minima at these points. The calculation therefore explicitly shows that the anti-skyrmions are not stable or metastable in this system, and again demonstrates the ability of metadynamics to explore the full landscape without prior knowledge or assumption about the state space. 

We now compare the free energy landscapes between the different temperatures in Figs.~\ref{fig:free_energy_surfaces}a,b,c. At the lowest temperature ($T=20$~K), the minimum energy path between the $\mathrm{FM}_\pm$ and $\mathrm{Sk}_{1\mp}$ states is deep and narrow. With increasing temperature, all of the features of the free energy surface broaden. Minima become shallower and smeared across the $Q$ axis, showing how thermal fluctuations cause fluctuations in the topological charge. The skyrmions become less like circular particles and become ragged due to the spin fluctuations, taking us away from the idealized concept of a smooth texture that covers the sphere. The upper and lower edges of $m_z = \pm 1$, in red, become thicker as the saturation magnetization decreases with temperature. The location of the energy minima move correspondingly towards smaller values of $|m_z|$. 

We extracted the minimum energy paths from the free energy surface using the \textsc{MEPSAnd} software~\cite{MarcosAlcalde_Bioinformatics_36_956_2020, *MEPSAnd_software}. The minimum-energy paths between the $\mathrm{FM}_\pm$ and $\mathrm{Sk}_{1\mp}$ states are marked with red dotted lines on top of Figs.~\ref{fig:free_energy_surfaces}b and c. At temperatures $T < 37$~K, the lowest energy path to destroy a skyrmion is always through the `collapse' process, where the skyrmion skrinks to a single spin which then flips. On the energy surface, this appears as an increase in $m_z$ due to shrinkage, followed by a change in $Q$ as the covering of the sphere reduces. Above $T=36$~K the minimum energy path changes, and the skyrmion is destroyed through a `singularity' process. A Bloch point forms on the edge of the skyrimon, causing it to lose topological charge and become a trivial domain, which can then relax to the uniform magnetic state. In this case, we can see the path on the energy surface first as a change in the topological charge followed by an increase in magnetization. 

\begin{figure}
    \centering
    \includegraphics[width=0.48\textwidth]{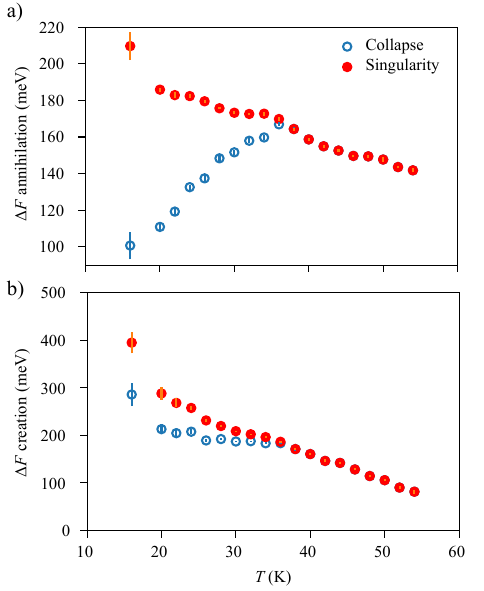}
    \caption{Free energy barrier for (a) skyrmion annihilation ($\mathrm{Sk}_{1\mp} \rightarrow \mathrm{FM}_\pm$) and (b) skyrmion creation ($\mathrm{FM}_\pm \rightarrow \mathrm{Sk}_{1\mp}$), along the collapse (blue open circles) and singularity (red filled circles) paths. Vertical lines show the standard error from calculating the energy barrier across 10 independent simulations.}
    \label{fig:energy_barriers}
\end{figure}

In Figs.~\ref{fig:energy_barriers}a and b we plot the height of the free energy barrier for skyrmion creation and annihilation, for both the collapse and singularity paths, as a function of temperature. The energy barrier is asymmetric because the ferromagnetic state is lower in energy than the skyrmion state; Creating a skyrmion costs more energy than destroying a skyrmion. At low temperature ($15$~K) the energy barrier to destroy a skyrmion via the singularity path is twice that of the collapse path, in agreement with the zero temperature results of the nudged elastic band method~\cite{CortesOrtuno_SciRep_7_4060_2017}. However, as the temperature increases, the singularity energy barrier decreases, whereas the collapse energy barrier increases. The temperature dependence is linear for both barriers, but the temperature dependence of the increase in the height of the collapse barrier is stronger. At a certain temperature, $T=37$~K, there is a crossover, and the singularity path becomes the lowest energy path.

The increasing energy barrier with temperature for the collapse process seems counter-intuitive because the anisotropy, exchange stiffness, and DMI decrease with increasing temperature due to spin fluctuations~\cite{Rozsa_PhysRevB_96_094436_2017, Tomasello_PhysRevB_97_060402_2018}. Breaking the skyrmion texture would appear to be an increasingly easy when considering only the internal energy. Indeed, the energy barrier from $\mathrm{FM}_+$ to $DW$ decreases with increasing temperature. However, we must also consider the entropy bottleneck of the process, having to transition through a highly specific spin state with a single spin flip. The thermal spin fluctations introduce entropy, which must be worked against to cause the shrinking and eventual collapse. This is similar to the high free energy cost of trying to force $m_z$ to be larger than its thermodynamic equilibrium value. Our results therefore show that the entropy has a more significant effect on the collapse free energy barrier than the renormalization of the magnetic parameters. This supports the conclusions that have been drawn on the role of applied magnetic fields on skyrmion lifetimes~\cite{Wild_SciAdv_3_2017}.

The energy barrier to annihilate a skyrmion via the singularity process decreases with temperature. Here, entropy is less important because any single spin on the circumference of the skyrmion that suffers a large enough fluctuation can trigger the process. The larger the thermal spin fluctuations, the more likely this will occur. Therefore, the spin fluctuations and magnetic parameter renormalization combine to cause the energy barrier to decrease with increasing temperature.

The change in the energy barriers to create a skyrmion (Fig.~\ref{fig:energy_barriers}b) with temperature is different. For both paths, the barrier decreases with increasing temperature. The collapse path is almost flat, while the singularity path reduces significantly. They cross at the same temperature as the annihilation barriers. The thermal spin fluctuations decrease the energy needed to form Bloch points, providing some solid angle between neighboring spins either to start the expansion process or on the edge of a domain wall to follow the singularity path. The increasing entropy term $TS$ is always helpful in this process.

\textit{Conclusions} -- Our results show that thermal spin fluctuations cannot be neglected in the calculation of energy barriers for skyrmion creation and annihilation. Spin fluctuations can, in fact, lead to changes in the minimum energy path, depending on the ambient temperature. At higher temperatures, the singularity process appears to be lower in energy, allowing the creation of skyrmions to form starting with topologically trivial bubble domains. This may explain why skyrmions can be generated using simple field cycling protocols~\cite{Zeissler_NatNanotechnol_13_1161_2018}. The applied field will alter the landscape, pushing the system along the $m_z$ axis, making it easier to fall into the skyrmion basins. The increase in the collapse energy barrier highlights that entropy plays an important role in skyrmion creation and annihilation. Overall, our work also demonstrates the usefulness of metadynamics in studying spin textures in magnetic materials at finite temperatures, with the method being easy to adapt to study many different systems through the careful choice of appropriate collective variables.

\section{Data Access}
Data is available upon reasonable request. The Monte Carlo spin modeling software used to generate the data (\textsc{JAMS}) is not currently open source while intellectual property issues are being resolved, but can be made available to individual researchers upon request. 

\section{Author Contributions}
\textit{Ioannis Charalampidis}: methodology, investigation, software, writing - review and editing. \textit{Joseph Barker}: conceptualization, methodology, software, writing - original draft, funding acquisition.

\section{Acknowledgments}

The authors thank Thomas Nussle for useful discussions during this work. J.B. acknowledges support from the Royal Society through a University Research Fellowship. I.C. acknowledges support through an EPSRC Doctoral Training Partnership. Calculations were performed on ARC4, part of the High Performance Computing facilities at the University of Leeds.

\bibliography{biblio}

\end{document}

\title{Supplementary Information:Metadynamics calculations of the effect of thermal spin fluctuations on skyrmion stability}

% \author{Ioannis Charalampidis}
\author{Ioannis Charalampidis\,\orcidlink{0009-0007-1174-5883}}
\email{ioannis.charalampidis@manchester.ac.uk}
\affiliation{School of Physics and Astronomy, University of Leeds, Leeds LS2 9JT, United Kingdom}

% \author{Joseph Barker}
\author{Joseph Barker\,\orcidlink{0000-0003-4843-5516}}
\email{j.barker@leeds.ac.uk}
\affiliation{School of Physics and Astronomy, University of Leeds, Leeds LS2 9JT, United Kingdom}

\maketitle

\section{S1: Calculation of topological charge}

We use the definition of the topological charge
\begin{equation}
    Q = \frac{1}{4\pi} \int \rmd^2 x ~\vec{S}\cdot \left(\frac{\partial \vec{S}}{\partial x} \times \frac{\partial \vec{S}}{\partial y}\right),
\end{equation}
which we calculate numerically on the lattice using finite differences for the gradients
\begin{align}
\begin{split}
        \frac{\partial \vec{S}(\vec{r}_i)}{\partial x} &\approx \frac{\Delta \vec{S}(\vec{r}_i)}{\Delta x} = \frac{1}{6}\left[2\vec{S}(\vec{r}_i+\vec{a}) - 2\vec{S}(\vec{r}_i-\vec{a}) + 
    \vec{S}(\vec{r}_i + \vec{a}+\vec{b}) + \vec{S}(\vec{r}_i - \vec{b}) - \vec{S}(\vec{r}_i - \vec{a}-\vec{b}) - \vec{S}(\vec{r}_i + \vec{b})\right]\\
    \frac{\partial \vec{S}(\vec{r}_i)}{\partial y} &\approx  \frac{\Delta \vec{S}(\vec{r}_i)}{\Delta y} = \frac{\sqrt{3}}{6}\left[\vec{S}(\vec{r}_i+\vec{b}) + \vec{S}(\vec{r}_i+\vec{a}+\vec{b}) - \vec{S}(\vec{r}_i-\vec{a})- \vec{S}(\vec{r}_i-\vec{a}-\vec{b}) \right]
\end{split}
\end{align}
The topological charge is then a sum over the finite differences
\begin{equation}
    Q = \frac{1}{4\pi}\sum_i \vec{S}(\vec{r}_i)\cdot\left(\frac{\Delta \vec{S}(\vec{r}_i)}{\Delta x}  \times \frac{\Delta \vec{S}(\vec{r}_i)}{\Delta y}  \right).
    \label{eq:topological_charge_finite_diff}
\end{equation}

\section{S2: Additional details of the metadynamics implementation}

For computational efficiency (avoiding many exponential function calls), we discretize our metadynamics potential on a grid with $\Delta Q = 0.05$ and $\Delta m_z = 0.01$. To find the value of the potential at an arbitrary point $V(Q, m_{z})$ we use bilinear interpolation from the nearest grid points. 

\section{S2: Harmonic Spring Boundary Conditions}

To avoid the system from exploring ever higher values of $Q$ we follow Refs.~\onlinecite{Laio_RepProgPhys_71_126601_2008, Laio_JHighEnergyPhys_2016_89_2016} and apply a harmonic spring potential outside of a region of interest. The references are solving for forces in the system, but we can interpret the forces as spring-like and write the appropriate potential energy. When the collective variable $Q$ is outside of this region, $|Q| > Q_{\mathrm{max}}$ the potential used is
\begin{equation}
    V(Q, m_z) = k(Q - Q_{\mathrm{max}})^2 + V(Q_{\mathrm{max}}, m_z) \quad \mathrm{when} \quad |Q| > Q_{\mathrm{max}}
\end{equation}
where $k$ is a spring constant with units of energy.  $V(Q_{\mathrm{max}}, m_z)$ is the value of the potential at the boundary which has built up due to the metadynamics Gaussians. It must be added to the spring potential outside of the region of interest to ensure that the total potential is smooth and continuous at the boundary. We stop adding new Gaussians to the landscape while $|Q| > Q_{\mathrm{max}}$. This boundary condition causes the system to be smoothly pushed back into the region of interest. In this work, we used $k = 0.05$~meV and $Q_{\mathrm{max}}=2.5$.

\bibliography{biblio}